\documentclass[12pt]{article}

\usepackage{array} 
\usepackage{epsfig}
\usepackage{amssymb}
\usepackage{amsmath}                          
\usepackage{graphics,graphpap}
\usepackage{graphicx}
\usepackage{epstopdf}

\renewcommand{\thefootnote}{\fnsymbol{footnote}}

\newcommand{\unit}[1]{\ensuremath{\rm\,#1}}
\newcommand{\invfb}{\unit{fb^{-1}}}
\newcommand{\fs}{\unit{fs}}

\newcommand{\invps}{\unit{ps^{-1}}}

\newcommand{\particle}[1]{{\ensuremath{\rm #1}}}
\newcommand{\BAR}[1]{\overline{#1}}
\newcommand{\MeVcc}{\unit{MeV\!/\!{\it c}^2}}
\newcommand{\Bs}{\particle{B_s}}
\newcommand{\Bsbar}{\particle{\BAR{B}{_s}}}

\newcommand{\DGs}{\ensuremath{\Delta\Gamma_{\rm s}}}
\newcommand{\Gs}{\ensuremath{\Gamma_{\rm s}}}
\newcommand{\DGsGs}{\ensuremath{\Delta\Gamma_{\rm s}/\Gamma_{\rm s}}}
\newcommand{\dms}{\ensuremath{\Delta m_{\rm s}}}

\newcommand{\Lg}[1]{L_{c;#1}^{g\phi}}
\newcommand{\Gev}{\,\rm GeV}
\newcommand{\Mev}{\,\rm MeV}

\begin{document}

\begin{titlepage}
\begin{flushright}\begin{tabular}{l}
IPPP/08/04\\
DCPT/08/08
\end{tabular}
\end{flushright}
\vskip1.5cm
\begin{center}
   {\Large \bf \boldmath Exploiting the width difference in $B_s \to \phi \gamma$}
    \vskip1.3cm {\sc
    Franz Muheim\footnote{muheim@ph.ed.ac.uk}
     and Yuehong Xie\footnote{yxie@ph.ed.ac.uk}
\vskip0.3cm    
 {\em  University of Edinburgh, Edinburgh EH9 3JZ, United Kingdom }
    \vskip0.5cm 
Roman Zwicky \footnote{Roman.Zwicky@durham.ac.uk}
  \vskip0.3cm
        {\em IPPP, Department of Physics, 
University of Durham, Durham DH1 3LE, UK}} \\
\vskip1.5cm

\vskip2.5cm

{\large\bf Abstract:\\[10pt]} \parbox[t]{\textwidth}{
The photon polarization in $B \to V \gamma$ is a sensitive
probe of right-handed currents.  In the
time dependent decay rate of $B_s \to \phi \gamma$ 
the coefficients $S$ and $H$ in front of the $\sin(\Delta m_s t)$ and
the $\sinh(\Delta \Gamma_s /2  t)$ terms are sensitive to those 
right-handed currents.
As compared to the $B_d$ system there is a sizable width difference   
in $B_s$ mesons which leads to the additional  measurable observable $H$. 
We show with a Monte Carlo simulation that the expected resolution  on 
$S$ and $H$ will be about $0.15$  at the LHCb experiment 
for $\Delta \Gamma_s/\Gamma_s = 0.15$ and a data sample of 
$2\, {\rm fb}^{-1}$. 
We also show that the observable $H$ can be 
measured from the untagged decay rate of $B_s$ mesons 
which has considerable experimental advantages as 
no flavour tag will be required.
The resolution on $H$ is inversely proportional to
the $B_s$ width difference $\Delta \Gamma_s$. 
These experimental prospects have to be compared with
the Standard Model predictions $S_{\phi \gamma} = 0\pm 0.002 $ and 
$H_{\phi \gamma} = 0.047 \pm0.025+0.015$ presented in this paper.
We also give  the Standard Model prediction and the experimental sensitivity 
for the direct CP asymmetry in $B_s \to \phi \gamma$.
}

\vfill

\end{center}
\end{titlepage}

\setcounter{footnote}{0}
\renewcommand{\thefootnote}{\arabic{footnote}}

\newpage

\section{Introduction}
Flavour changing neutral current (FCNC) decays are forbidden 
at tree level in the Standard Model (SM) and are therefore a sensitive probe of new physics (NP).
Furthermore, the $V\!-\!A$ structure of the weak interactions can be
tested in  FCNC decays of the type $b \to (d,s) \gamma$, since
the emitted photon is predominantly left-handed. 
The crucial point is that  the weak force only couples 
to left-handed quarks. The structure of the leading operator 
$Q_7 \sim \bar s \sigma_{\mu\nu} F^{\mu\nu} b_{L(R)}$ necessitates
a helicity flip on the external quark legs, 
which introduces a natural hierarchy between the left and
right-handed production of the order of $m_{d,s}/m_b$.
However, it is difficult to measure the helicity of the photon directly, e.g. \cite{measure_pol}.
It was pointed out ten years ago
that the time dependent CP asymmetry is  an indirect measure of the photon helicity  \cite{AGS97}, since it is caused by 
the interference of the left  and right-handed helicity amplitudes.

At the B factories the exclusive radiative decays of the $B_d$ meson were studied. 
The coefficent $S$ in front of the $\sin(\Delta m_d t)$ term
in the time dependent CP asymmetry has been measured in 
$B_d \to K^{0*}(K_S\pi^0)\gamma$ at  the B factories 
BaBar $S_{K^*\gamma} =  -0.08\pm 0.31 \pm 0.05$
 \cite{babarBKg}  and Belle 
 $S_{K^*\gamma} = -0.32^{ +0.36}_{-0.33} \pm 0.05$ \cite{belleBKg}.  
 The average is $S_{K^*\gamma}^{\rm HFAG} = -0.19 \pm 0.23$ \cite{HFAG}.
 Recently  Belle reported a measurement of $S_{\rho \gamma} = -0.83\pm 0.65 \pm 0.18$ \cite{belleBrg} 
in $B \to \rho^0 \gamma$.
 Comparing the experimental values with 
 theoretical predictions \cite{AGS97}  \cite{charmloops} \cite{BJZ}
 it is clear that larger data samples are required before 
conclusions can be drawn.

The large production rate of $B_s$ mesons at the LHC
opens up the possibility to study the  $B_s$ system with high statistical
precision. 
In this letter we intend to argue that the $B_s \to \phi \gamma$ decay
is a particularly promising channel 
to test the $V\!-\!A$ structure of the SM at the LHCb experiment. 
This method is independent of the actual value of the $B_s$ mixing angle,
since there is a measurable coefficient in front of the
$\sinh \big( \Delta \Gamma_s /2 \, t \big)$ term
in the time dependent decay rate, 
which we shall denote by the letter $H$.

At the level of the QCD calculation the decay $B_s \to \phi \gamma$ 
is very similar to $B_d \to K^* \gamma$. Compared to the $B_d$ meson,
the new elements of the $B_s$ meson are the small mixing phase $\phi_s$ and the large width difference $\Delta \Gamma_s$ of the $B_s$ meson, which will play a central role
in this letter.
The SM predictions for the mixing angles and widths are 
\begin{alignat}{2}
\label{eq:miximaxi}
\phi_{s} & \simeq  -2\lambda^2 \eta \simeq -2^\circ 
\qquad 
& \phi_{d} &  \simeq    2 \beta \simeq 43^\circ  \nonumber \\
 \frac{\Delta \Gamma_s}{\Gamma_s}    &= 0.107\pm 0.065
\qquad   & \frac{\Delta \Gamma_d}{\Gamma_d}&=   
\big(40.9^{+8.8}_{-9.9} \big) \cdot 10^{-4} \, ,
\end{alignat}
where the values of the widths are taken from the recent update
of Ref.~\cite{LenzNierste}. The Wolfenstein  parameters
$\lambda \simeq 0.227(1) $ and $\eta \simeq 0.34(4)$ are taken from \cite{PDG}.
While the width and phase of the $B_d$ meson are precisely measured and 
consistent with the SM within uncertainties \cite{PDG}, the knowledge of
 the $B_s$ width and the mixing phase is still  poor. 
The D0 experiment finds 
 $ \phi_s = -0.70 ^{+0.47}_{-0.39}$  and $\DGs = 0.13 \pm 0.09 \invps$ \cite{Asl_Bs}. 
Combining this result of $\DGs$ with other measurements, the Heavy Flavour Averaging Group quotes
$\DGs = 0.071 ^{+0.053}_{-0.057}$ and  $\frac{\Delta \Gamma_s}{\Gamma_s} = 0.104 ^{+0.076}_{-0.064}$ 
\cite{HFAG}.

In this paper we will show that the experimental resolution is independent 
of the actual value for the coefficients $S$ and $H$ of the $\sin(\Delta m t)$ and 
$\sinh(\Delta \Gamma/2 t)$ terms.
Therefore it is crucial that either $S$ or $H$ is sizable
in order to detect NP from enhanced right-handed currents
as opposed to NP in the mixing\footnote{The mixing angle $\phi_s$ itself will be measured 
in a clean way at the LHCb experiment through the CP asymmetry 
$S_{B_s \to J/\Psi \phi} \sim \sin(\phi_s)$ in the decay
$B_s \to J/\Psi \phi$.}.
In the SM the short distance contribution dominates which 
has a single weak phase which is exactly cancelled by the
mixing phase. 
Since $S$ and $H$ are proportional to the sine and cosine
it is more likely that NP will be sizable in $H$ rather than $S$.
In the $B_d$ system only $S$ is measurable, since the
width is too small, but fortunately $S$ is sizable 
because the phases from the mixing and the short distance 
process do not cancel.
We refer the reader to appendix \ref{app:formulas} 
for  formulae on $S$ and $H$ in terms of two weak amplitudes
which go beyond the simplified discussion in this introduction.

The paper is organised as follows.
Definitions of the observables and theory predictions
including the non-local charm loop contribution \cite{charmloops}
are  presented in section \ref{sec:theory}.
Further  useful formulae are compiled in the appendix 
\ref{app:formulas}.
The extraction of the  observables from the time dependent decay rates is discussed in section \ref{sec:extraction} and  a Monte Carlo
simulation for the experimental accuracy is presented in section 
\ref{sec:montecarlo}. The letter ends with  conclusions 
in section \ref{sec:conclusions}.

\section{Time dependent CP-violation in $B_s \to \phi \gamma$}
\label{sec:theory}

The normalised CP asymmetry, for $B_s \to \phi \gamma$ 
is defined as follows
\begin{equation}
\label{eq:CPdef}
{\cal A}_{\rm CP}(B_s \to \phi \gamma) \equiv \frac{\Gamma[\bar B_s \to  \phi \gamma] -\Gamma[ B_s \to  \phi \gamma]}{\Gamma[\bar B_s \to \phi \gamma] + 
\Gamma[ B_s \to  \phi \gamma]},   
\end{equation}
where the left and right-handed photon contribution are added
incoherently 
$\Gamma[ B_s \to  \phi \gamma] = \Gamma[ B_s \to  \phi \gamma_L]
+ \Gamma[ B_s \to  \phi \gamma_R]$.
Neutral mesons, such as the $B_s$, exhibit a time dependence in 
the CP asymmetry through mixing, if the particle and the antiparticle allow
for a common final state. 
In $B_s \to \phi \gamma$ this amounts to
\begin{equation}
\label{eq:both}
B_s \rightarrow  \phi \gamma_{L(R)}  \leftarrow \bar B_s \,.
\end{equation}
 The general time evolution
of the decay rates parameterised in terms of the 
amplitudes  can be found in \cite{PDG}. The ratio of coefficients $p$ and $q$
\begin{equation}
\Big(\frac{q}{p}\Big)_s = \Big|\frac{q}{p} \Big|_s e^{-i\phi_s} \quad ,
\end{equation}
relating the physical and the flavour eigenstates, characterizes 
the mixing of the $B_s$ mesons. 
The $B_s$ mixing phase $\phi_s$ is small when compared to the mixing phase 
in $B_d$ mesons, c.f. Eq.~\eqref{eq:miximaxi}.
The absolute value of $(q/p)_s$ can be determined experimentally from the semileptonic CP asymmetry. 
The measurement of the latter  \cite{Asl_Bs} indicates that the quantity is very close to unity, 
$1-|q/p |_{s} =  (0.05 \pm 0.45) \cdot 10^{-3} $.

With $|q/p|_s=1$ the CP asymmetry \eqref{eq:CPdef} assumes
the following generic time dependent form\footnote{In the literature the notation
$C = -{\cal A}_{\rm dir}$, $S = {\cal A}_{\rm mix}$ and $H = \pm {\cal A}_{\Delta \Gamma}$
is frequently used.}
\begin{equation}
\label{eq:CPform}
{\cal A}_{\rm CP}(B_s \to \phi \gamma)[t]
= \frac{S \sin(\Delta m_s t) -C \cos(\Delta m_s t)}
{ {\rm cosh}(\frac{\Delta \Gamma_s}{2}t) - H  {\rm sinh}(\frac{\Delta \Gamma_s}{2} t) } \, .
\end{equation}
The mass difference and the width difference are defined as
$\Delta m_s = m_H -m_L > 0 $  $\Delta \Gamma_s = \Gamma_L - \Gamma_H$,  where the subscripts $H$ and $L$ stand
for heavy and light respectively. The definition of the width difference corresponds to a positive value  in
the SM, i.e. $\Delta \Gamma_s^{SM} > 0$.
In terms of the left-handed and right-handed amplitudes,
\begin{equation}
\label{eq:para}
{\cal A}_{L(R)} \equiv {\cal A}(B_s \to \phi \gamma_{L(R)} ) \qquad 
\bar {\cal A}_{L(R)} \equiv {\cal A}(\bar B_s \to \phi \gamma_{L(R)} ) \,,
\end{equation}
the observables $C$, $S$ and $H$ assume the following form
\begin{eqnarray}
\label{eq:CPamp}
C &=&   \frac{(|{\cal A}_L|^2 + | {\cal A}_R|^2)
-  (|\bar {\cal A}_R|^2 +  |\bar {\cal A}_L|^2 )}
{ |{\cal A}_L|^2 + |\bar {\cal A}_L|^2 
+  |{\cal A}_R|^2 + |\bar {\cal A}_R|^2}  \nonumber \\
S &=&  \frac{2  \, {\rm Im}[\frac{q}{p}(\bar {\cal A}_L {\cal A}_L^* +
\bar {\cal A}_R {\cal A}_R^*)]}
 { |{\cal A}_L|^2 + |\bar {\cal A}_L|^2 
+  |{\cal A}_R|^2 + |\bar {\cal A}_R|^2}  \nonumber \\
H &=&  \frac{2 \, {\rm Re}[\frac{q}{p}(\bar{\cal A}_L
 {\cal A}_L^* + \bar {\cal A}_R {\cal A}_R^*)]}
 { |{\cal A}_L|^2 + |\bar {\cal A}_L|^2 
+  |{\cal A}_R|^2 + |\bar {\cal A}_R|^2 } \, .
 \end{eqnarray}
 The amplitudes are parametrised in terms of the CKM 
 phases according to   Eq.~\eqref{eq:amp_dec} in the appendix,
although with a different normalisation,
\begin{equation}
\label{eq:dec2}
{\cal A}_{L(R)} =  \frac{G_F}{\sqrt{2}} \Big[-\frac{e\, m_b}{2 \pi^2}\, T_1^{B_s \to \phi}(0)\Big] \left( \lambda_u { a}^u_{L(R)} 
+  \lambda_c { a}^c_{L(R)}+  \lambda_t { a}^t_{L(R)} \right)S_{L(R)}   
\, ,
\end{equation}
where $m_b$ is the $b$ quark mass, $G_F$ is the Fermi constant,
$\lambda_U = V_{\rm Us}^* V_{\rm Ub}$ are CKM factors  with $U = \{u,c,t\}$ and
$T_1(0) = 0.31(4)$ is a penguin form factor \cite{BZ04b} whose value 
was updated in \cite{BJZ}. 
The left-right projectors are
\begin{equation}
S_{L(R)} = \epsilon^{\mu\nu\rho\sigma} e_\mu^* \eta_\nu^* p_\rho q_\sigma \pm i
\{ (e^* \eta^*) (pq) - (e^*p)(\eta^* q)\} \,,
\end{equation}
where $e_\mu(q)$  and $\eta_\nu(p)$ are the photon and $\phi$ polarisation vectors and $q$ and $p$ are the photon
and $\phi$ four-momentum, respectively. 
The decomposition in Eq.~\eqref{eq:dec2} is ambigous since 
the three generation unitarity $\lambda_u + \lambda_c + \lambda_t = 0$ 
allows us  to reshuffle terms from one amplitude into the other.
Often it is  convenient to eliminate one amplitude
by invoking the unitarity relation, e.g. formulae in the appendix \ref{app:formulas}.
For notational clarity we shall quote,
\begin{equation}
\label{eq:Heff}
H_{\rm eff}=\frac{G_F}{\sqrt{2}}\left( \sum_{U=u,c}\lambda_U 
 (C_1 Q^U_1 + C_2 Q^U_2) + \lambda_t \sum_{i=3\ldots 8} C_i Q_i\right) \,,
\end{equation}
the total $b \to s \gamma$ effective  Hamiltonian.
In the SM the leading operator,
\begin{equation}
\label{Q7}
Q_7 = \frac{e}{8\pi^2} \left[ m_b \bar s \sigma_{\mu\nu} (1\!+\! \gamma_5)
b + m_s \bar s \sigma_{\mu\nu} (1\!-\! \gamma_5) b\right] F^{\mu\nu} \,,
\end{equation}
is due to short distance penguin processes.
This leads to a particular chiral pattern \cite{AGS97} 
due to the $V\!-\!A$ structure of the weak interactions. 
Namely, the $\bar B_s(B_s)$ meson decays predominantly into 
a left(right)-handed photon whereas the decay of  the $B_s(\bar B_s)$ meson  
into the left(right)-handed photon is suppressed by a $m_s/m_b$ chirality 
factor,
 \begin{equation}
 \label{eq:at}
\left( \begin{array}{l}
a_L^t     \\
a_R^t  
\end{array} \right)
 = C_7  \, 
\left( \begin{array}{l}
1 \\
m_s/m_b
\end{array} \right) + O(1/m_b,\alpha_s) \quad .
\end{equation} 
Due to the interference of mixing and decay in $B_s \to \phi \gamma$,
a single weak decay amplitude proportional to $\lambda_t$ is
exactly cancelled by the mixing phase,
$$
H_{Q_7} =  2 \frac{m_s}{m_b} \cos(\phi_s\!-\! \phi_s) = 2 \frac{m_s}{m_b} \qquad  
S_{Q_7} = - 2 \frac{m_s}{m_b} \sin(\phi_s\! -\! \phi_s) = 0  \, .
$$
Note that at this stage the CP asymmetry pattern is analogous  to 
$B_s \to J/\Psi \phi$ up to the chiral suppression of the interference term.
The formula for $S$ was presented in the original paper \cite{AGS97}. 
Later it was pointed out by Grinstein et al \cite{alt} that QCD alters the $V\!-\!A$ pattern and that the current operator $Q_2^U$,
\begin{equation}
\label{eq:Q2U}
Q_2^U = \bar s\gamma_\mu (1\!-\!\gamma_5) U \, \bar
U \gamma^\mu(1\!-\! \gamma_5)b \qquad U = \{u,c\} \,,
\end{equation}
might lead to sizable corrections in part due to its large Wilson coefficient
$|C_2| \simeq 3 |C_7|$. 
The dominant contribution corresponds to the physical process of
emission of a collinear gluon from the long distance charm loop into the vector meson final state. 
In reference \cite{cpas} the charm loop was expanded to leading order 
in $1/m_c^2$, for which a large uncertainty was attributed, 
and the remaining matrix element was estimated with 
Light-Cone Sum Rules (LCSR)\footnote{The same expansion and local QCD sum rules were also used in reference \cite{KRSW97} in the conjunction with the total branching fraction.}.
The contribution turned out to be relatively small, suppressed by large loop factors. 
In reference \cite{charmloops} the charm loop is calculated  to all orders in $m_c$ within the framework of the light-cone expansion. 
The closeness to the charm threshold results in a large strong phase.
The expansion in the charm mass does not reveal the phase and is not convergent when higher orders are taken into account.
Nevertheless the first order and the all order result differ by less 
than a factor of two which is well within the uncertainty attributed 
in \cite{cpas}.
The numerical result is \cite{charmloops}
\begin{equation}
\left( \begin{array}{l}
a_L^c     \\
a_R^c  
\end{array} \right)
 =  C_2 \frac{Q_c}{T_1^{B_s \to \phi}(0)}  \, 
\left( \begin{array}{l}
 \Lg{L}(0)  \\
 \Lg{R}(0) 
\end{array} \right)  \qquad Q_c = \frac{2}{3} \,,
\end{equation} 
where $Q_c$ is the charge of the charm quark and 
\begin{eqnarray}
\label{eq:Lc}
 \Lg{L}(0)  =     (4.8 \cdot 10^{-3}  \pm 70\%) e^{i (255 \pm 15)^\circ }     \quad 
 \Lg{R}(0)  =    (1.8 \cdot 10^{-3}  \pm 70\%) e^{i (106 \pm 15)^\circ } \,.
\end{eqnarray}
Results for the up quark loops,  due to $Q_2^u$ \eqref{eq:Q2U},
can be found in reference \cite{BJZ}.
They are generally not sizable in $b \to s$ transitions because of the CKM hierarchy $|\lambda_u|  \ll |\lambda_c| \simeq |\lambda_t|$. 
The contributions in \eqref{eq:Lc}
will have a minor impact on the observables $S$ and $H$ 
because they are almost imaginary and the left-handed one is larger 
than the right-handed one.
It is therefore natural to ask whether these patterns will remain for  
contributions other than short distance, 
such as the emission of the gluon from the B meson or 
hard spectator interactions beyond the leading $1/m_b$ term.
In reference \cite{charmloops} the emission of a soft gluon from the $B$ meson
to the charm quark loop is studied. Using an analogous  
notation as above it is found that, $L^{gB}_{c;L} = 0.03(20) \cdot 10^{-3}$ and 
$L^{gB}_{c;R} = 0.4(3) \cdot 10^{-3}$. These contributions are real and
the left-right hierarchy appears to be inverted. We will take these contributions as an estimate of the uncertainty due to non-short distance contributions. 

We will now turn to the results 
of the parameters $S$ and $H$. We use the formulae given in the
 appendix in Eq.~\eqref{eq:twoamp} and obtain
\begin{equation}
\label{eq:inter}
H =   0.047 (1  \, \pm  17\%_{m_s}   \pm 10 \%_{LD} \pm 14\%_{\delta_{\Lg{R}}} \pm 5\%_{|\Lg{R}|}
)  \qquad 
S =  0  \pm  0.002  \, ,
 \end{equation}
where we have indicated parametric relative uncertainties for the strange quark mass
$m_s(2\Gev) = 100(20) \Mev$,  further long distance contributions mentioned above  
and for the collinear gluon  $\{|\Lg{R}|,\delta_{\Lg{R}}\}$ as given in
 Eq.~\eqref{eq:Lc}.
The uncertainty of the latter is small because the imaginary part does 
not contribute to the time dependent CP asymmetry when it interferes 
with the dominant and real $a_L^t$ in Eq.~\eqref{eq:at}.
In other words the strong phase difference is nearly ninety degrees
and gives a small contribution when the cosine is taken, c.f.
formula ~\eqref{eq:twoamp} in the appendix.
The leading contribution to the observable $S$ is given
by $2 {\rm Re}[a_R^{u*} / a_L^{t*}] |\lambda_u/\lambda_t| \sin(\gamma)$,
c.f.  using the notation in \eqref{eq:dec2} in the formula given in 
\eqref{eq:twoamp} in the appendix.
From this expression it is seen that $S$ is  CKM and helicity 
suppressed resulting in a vanishingly small value. 
For the uncertainty we assume that the  
helicity suppression of charm and up contributions is not larger 
than the one of the leading operator $Q_7$ \eqref{eq:at}. 
The uncertainty for $S$ and $H$ 
caused by the form factor $T_1$ and the Wilson coefficients
$C_2 = 1.03$ and $C_7 = -0.31$ are negligible due to cancellation in the
ratio.
 
Further uncertainties are coming from weak annihilation 
whose size does not contribute more than $5\%$ \cite{bosch,BJZ}
and contributions from the  gluon penguin operator $Q_8$,
where the gluon is emitted into the long distance photon 
wave function, are expected to be of the same size.
Hard spectator corrections to the chirality structure 
are of order $O(m_s/m_b)$ and, taking into account
the leading contribution from reference \cite{bosch}, are about $10\%$ if they should  contribute maximally to the right handed amplitude.
Another contribution comes from the gluon emission to the spectator quark 
which has been calculated in the perturbative QCD approach \cite{Sanda}
and indicates a shift of $\delta S_{K^*\gamma} = -0.01$ which we translate
into a one sided uncertainty for 
$\delta H_{\phi\gamma} = 0.015$ for $H_{\phi\gamma}$.
 Adopting a conservative estimate and adding the uncertainties 
in \eqref{eq:inter} linearly, another $10\%$ for the further 
contributions mentioned above and
the one sided spectator correction we 
arrive at our final estimate
\begin{equation}
\label{eq:result}
H_{\phi \gamma} =  0.047 \pm 0.025 + 0.015_{O(\alpha_s)} \qquad 
S_{\phi \gamma} = 0 \pm 0.002 \quad .
\end{equation}
Without the inclusion of  the charm loops the results is $H = 0.041$.
The result for $H$ is new whereas $S$ is almost the same as $-0.001(1)$ predicted in \cite{BJZ} up to the contribution of the charm loop 
which changes due to the large strong phase found in
\eqref{eq:Lc} as compared to the real values in \cite{BJZ}.
   
The CP asymmetry $C$ \eqref{eq:CPform} 
is sensitive to novel weak phases rather than to right handed currents.
It is proportional to the sine of the weak and 
strong phase and is given by
\begin{equation}
\label{eq:Cres}
C_{\phi \gamma} \simeq  - \underbrace{\frac{2 {\rm Im}[ \lambda_u^* \lambda_c]}{|
\lambda_t|^2}}_{2 \eta \lambda^2 \simeq 0.037}
\underbrace{ \frac{  {\rm Im} [ a_L^{u\,*} a_L^c ]} {C_7^2}}_{O(\alpha_s)}  \simeq  0.005(5) \quad ,
\end{equation}    
where we have used the notation given in Eq.\eqref{eq:dec2}
with $a_L^t$ eliminated by use of the three generation 
unitarity relation. Note that the right handed amplitudes are 
irrelevant since their contributions are of the order $O(m_s/m_b)$.
The numerically relevant imaginary parts are due to
charm loop contributions from the operator $Q_2^c$ \eqref{eq:Q2U}.
More specifically there are vertex corrections, hard spectator interactions and gluon emission into the final state.
The first two contributions are taken from
\cite{bosch} and the
gluon emission is given by $L^{g\phi}_{c;L}$ in Eq.\eqref{eq:Lc}
and contributes about one third to the asymmetry.
The CP asymmetry is small since it is 
CKM and $O(\alpha_s)$ suppressed. 
For the uncertainty in Eq.\eqref{eq:Cres} is due to the one
given in  Eq.\eqref{eq:Lc} for the emission of the gluon into the final state and an assumed a similar precision for the short distance and hard spectator contributions.

 After the theoretical prediction we will now turn in the next sections 
 to the experimental prospects for measuring the observables
 $S$, $H$  and $C$.

\section{Extraction of observables}
\label{sec:extraction}

The observables $S$, $H$ and $C$, appearing in the
time dependent CP asymmetry \eqref{eq:CPform},
can be extracted from the time dependent 
decay rates.
Without considering any 
experimental effects,
the time dependent decay rate, ${\cal B}(t)$,
of a $B_s$ meson, produced at 
$t=0$, is given by
\begin{equation}
\label{eq:b-t}
  {\cal B}(t)= {\cal B}_0 e^{-\Gamma_st}
    [{\rm cosh}(\frac{\Delta \Gamma_s}{2}t) -H {\rm sinh}(\frac{\Delta \Gamma_s}{2}t)
               +C\cos(\Delta m_s t) - S \sin(\Delta m_s t)]
\end{equation}
and the  decay rate, $\bar {\cal B}(t)$, of a $\bar{B}_s$ at $t=0$ is given by
\begin{equation}
\label{eq:bbar-t}
  \bar{\cal B}(t) =
{\cal B}_0 e^{-\Gamma_st}[{\rm cosh}(\frac{\Delta \Gamma_s}{2}t) -H {\rm sinh}(\frac{\Delta \Gamma_s}{2}t)
               -C\cos(\Delta m_s t) + S \sin(\Delta m_s t)].
\end{equation}
where ${\cal B}_0$ is the total decay rate.
It is the quantity $H\DGs$ which can be experimentally measured 
since $H\sinh(\frac{\DGs}{2} t) \approx H \DGs t/2$ for small $\frac{\DGs}{2} t$. 
Thus, the determination of $H$ requires that the 
$B_s$ width difference $\DGs$ be measured elsewhere. 
This can be achieved by the LHCb experiment which,
using the  $\Bs \rightarrow J/\psi \phi$ data sample, 
will be able to reach a statistical precision of $\pm 0.0092$
on $\DGsGs$, up to a sign-ambiguity~\cite{bib:lhcbDGs}.
Therefore, in this study we 
assume that $\DGs$ is precisely known. 
We only need to perform the study of the time dependent 
decay rates at one given value of $\DGsGs$ as 
the sensitivity on $H$ is inversely proportional to the width difference $\Delta \Gamma_s$.

While the determination of the coefficients $S$ and $C$ 
relies on the knowledge of the initial flavour of the $B_s$ mesons, 
the extraction of the observable $H$ does not 
require flavour tagging. The observable $H$ can
be measured from the untagged time dependent decay rate spectrum
(\ref{eq:b-t},\ref{eq:bbar-t}) which, from an experimental point of view, 
makes this a very promising method. 
In the next section we will
investigate prospects to measure these observables in future experiments.

\section{Experimental prospects}
\label{sec:montecarlo}

In the Standard Model, the CP-averaged branching ratio of $B_s \rightarrow \phi \gamma$ is predicted to be~\cite{BJZ}:  
\begin{equation}
{\cal B}(B_s \rightarrow \phi \gamma) = (39.4\pm 10.7 \pm 5.3) \times 10 ^{-6} \; .
\end{equation}
 The CDF Collaboration searched for this decay in $p\bar p$ collisions and set an upper limit of 
${\cal B}(B_s \rightarrow \phi \gamma) <1.9 \times 10^{-3}$~\cite{bib:cdf_report}
at the $95\%$ confidence level. 
Using a data sample of 23.6 fb$^{-1}$ recorded at the 
$\Upsilon(5S)$ resonance, which corresponds to 
about 2.6 millions of $B_s$ mesons, 
the Belle Collaboration recently reported a measurement of
${\cal B}(B_s \rightarrow \phi \gamma) = 
(5.7 ^{+1.8+1.2}_{-1.5-1.7} )\times 10^{-5}$ 
with a significance of 5.5$\sigma$~\cite{bib:belle_report}.
LHCb is a dedicated B physics experiment at the Large Hadron Collider, and is 
expected to start data taking in 2008~\cite{bib:lhcb_tdr}. 
A data sample of $\sim 2\; \rm fb^{-1}$, 
which the LHCb experiment expects to accumulate in a nominal year, 
corresponds to  about $7\times 10^{10}$ produced \Bs\ (\Bsbar) mesons 
whose decay products will be inside the LHCb detector acceptance. 
This copius production rate for \Bs\ mesons will open a window 
for the search of physics beyond the SM.

LHCb has performed a detailed Monte Carlo simulation 
to estimate the performance of the event reconstruction 
for the decay $B_s \rightarrow \phi \gamma$~\cite{bib:lhcb_2007-030}.
In a 2 \invfb\ data sample 
about 11500 signal events are expected to pass the Level 0 trigger\footnote{The high level trigger 
is not taken into account in this reference.}
and the event selection criteria 
with an  upper limit on background over signal ratio of
$B/S < 0.55$ at $90\%$ confidence level.
The \Bs\ mass resolution is about 70 \MeVcc.
The flavour of a \Bs\ (\Bsbar) meson at production can be inferred from
the decay products of the opposite side $b$ hadron or from the charge of the kaon accompanying the 
production 
of the signal \Bs\ (\Bsbar) meson. Using simulated events, 
this tagging procedure yields an efficiency of 60$\%$
and a wrong-tag fraction of 30$\%$ at the LHCb experiment. 
The proper decay time resolution is estimated to be about 80~\fs.
In this study we take into account the high level trigger efficiency and
conservatively assume a signal yield of 7700 signal events
from an integrated luminosity of 2 \invfb\  
and a background over signal ratio of $0.62$.

Based on these yields for 2 \invfb\ of data 
and the experimental resolutions,
a toy Monte Carlo approach is used to evaluate the statistical errors on 
$C$, $S$ and $H$. 
The distributions for the proper decay time, the reconstructed \Bs\ mass,
the cosine of the polar angle of the $K^+$ in the rest frame of the $\phi$ 
meson ($\cos\theta$) and the flavour tag  
are described by a probability density function (PDF).
In each toy experiment this PDF is used to generate and fit the data. 
Then this toy experiment is repeated many times to produce 
distributions for $C$, $S$ and $H$, from which the statistical precision
can be determined.

The signal PDF is modelled using the theoretical distribution
for each observable convoluted by the following detector effects:
the  \Bs\ mass resolution, the proper decay time resolution, 
the reconstruction efficiency as a function of  proper  decay time,
the tagging efficiency and the wrong-tag fraction.
A simple model is employed to describe the background PDF. 
We assume that the background is uniformly distributed
in the \Bs\ mass and in $\cos\theta$, 
and has an exponential proper decay time spectrum 
with an effective lifetime which is one third of the signal lifetime.
The detector effects and the  
background distributions are assumed to be 
precisely known.
The  theoretical signal distributions contain 
the following physical parameters: $C$, $S$, $H$, 
the \Bs\ average decay width $\Gs$,  
the \Bs\ width and mass difference $\DGs$ and $\dms$,
and the  \Bs\ mass $m_{\Bs}$. 
In the fit $C$, $S$ and $H$ are free 
parameters. All the other parameters are fixed to their 
input values which are given in Table~\ref{table:input}. 
A possible \Bs-\Bsbar\ production asymmetry and CP violation
in the \Bs\ mixing ($|q/p|_s \neq 1)$ are neglected in this study.

\begin{table}[bht]
\begin{center}
\caption{Input values of the physical \Bs\ observables, except $C$, $S$ and $H$.}
\label{table:input}
\vspace{0.3cm}
\begin{tabular}{@{}|c|c|c|c|}
\hline
 $\Gs$ & $\DGs$ & $\dms$ & $m_{\Bs}$ \\ \hline

 0.67 \invps\ & 0.1 \invps\       & 17.0 \invps\       &  5369 \MeVcc       \\ \hline
\end{tabular}
\end{center}
\end{table}


A number of 500 toy experiments are generated for each set of 
values for $C$, $S$ and $H$.
For a baseline scenario we set the parameters
 $C = 0$,
 $S = 0$,
 and $H = 0$,
which is close to the SM prediction. 
Using a maximum likelihood fit 
we then determine these parameters in each toy experiment.

\begin{figure}[b]
\begin{center}
\includegraphics[width=6.5cm]{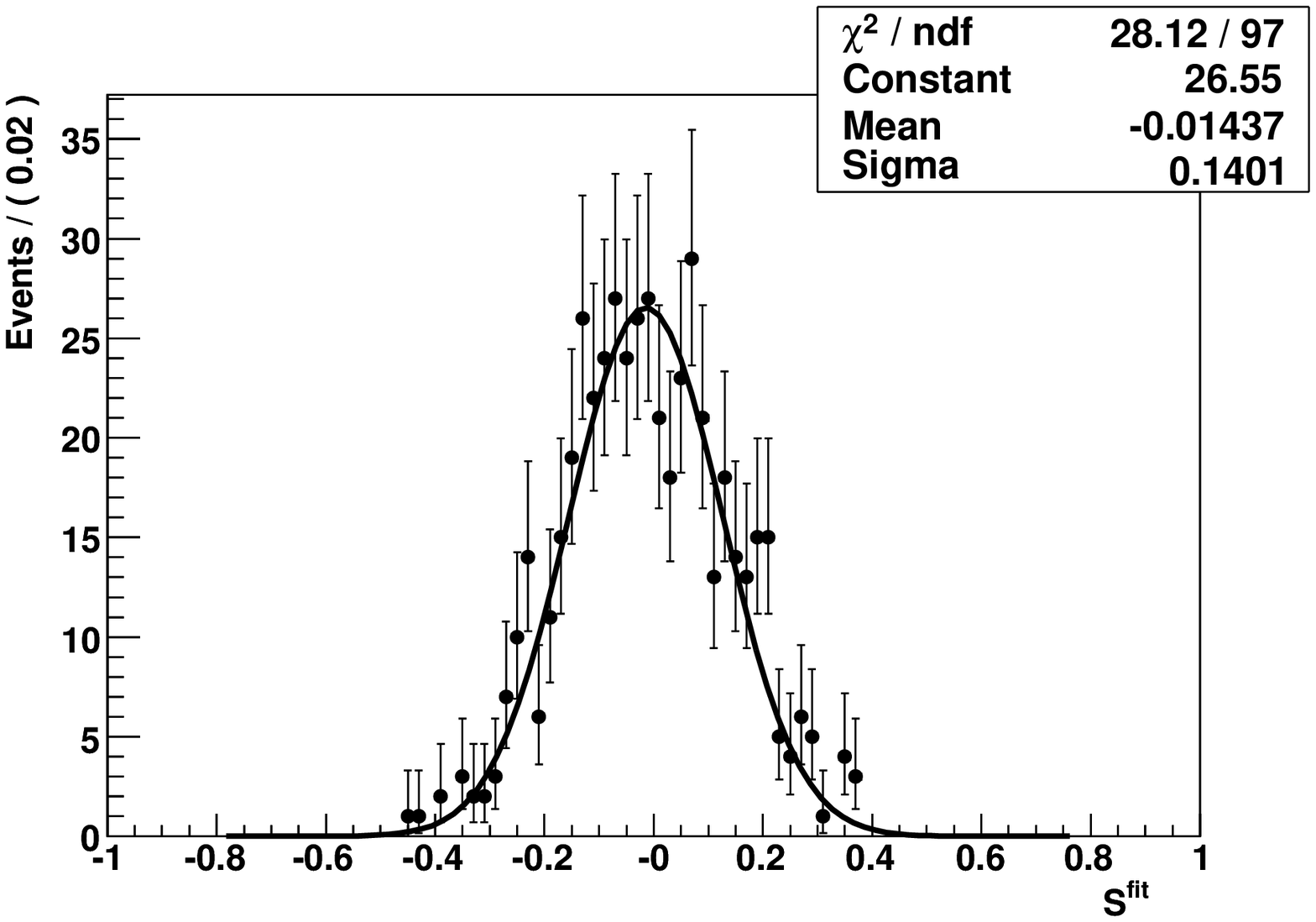}
\includegraphics[width=6.5cm]{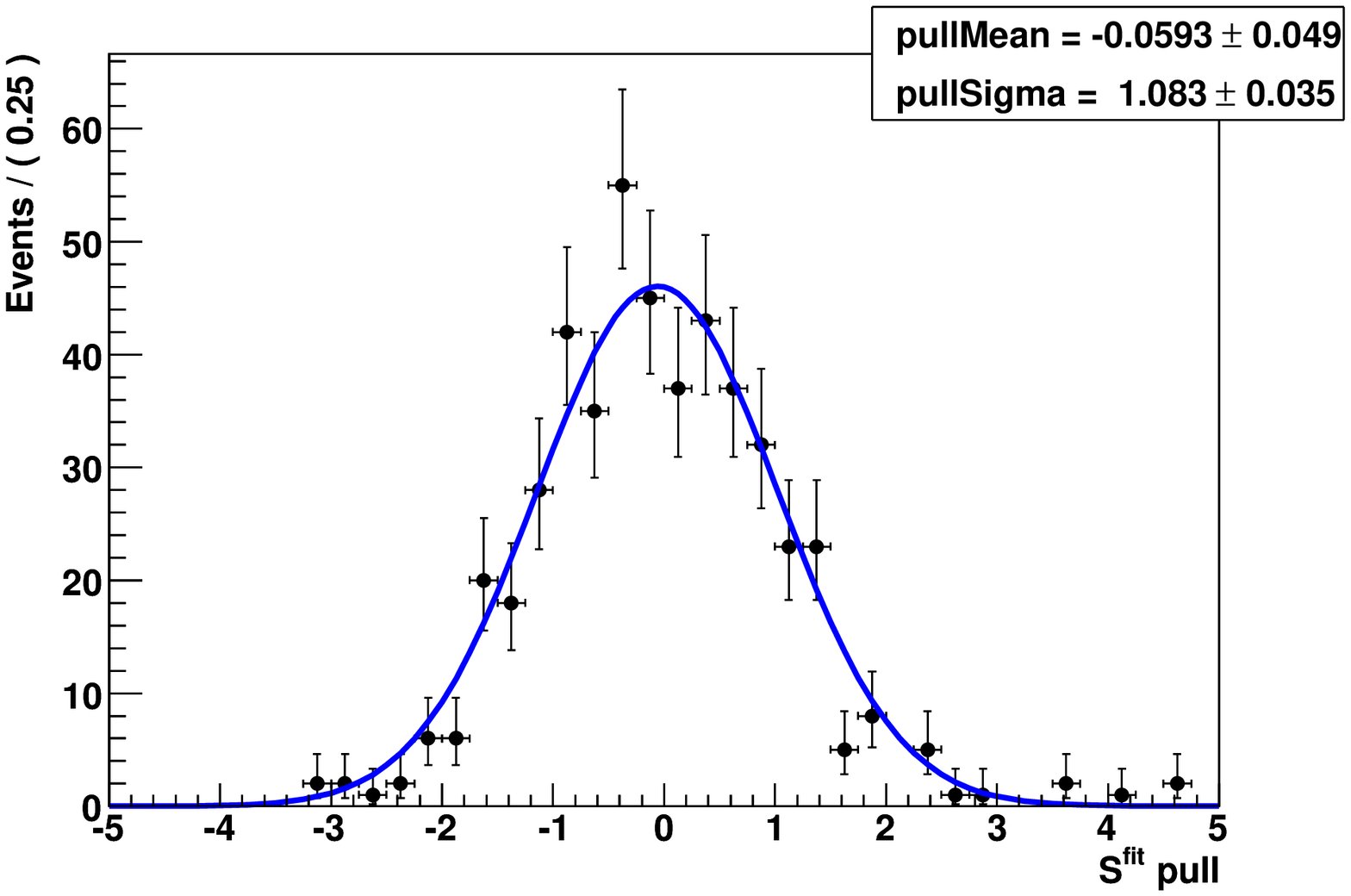}
 \caption{
  Left: the distribution of the fitted values $S^{fit}$ from 500 toy experiments
  for the baseline scenario with $S^{input} =0$ ;
  right: the distribution of $(S^{fit}-S^{input})/\epsilon_S$.
 }
\label{fig:sreso}
\end{center}
\end{figure}

\begin{figure}
\begin{center}
\includegraphics[width=6.5cm]{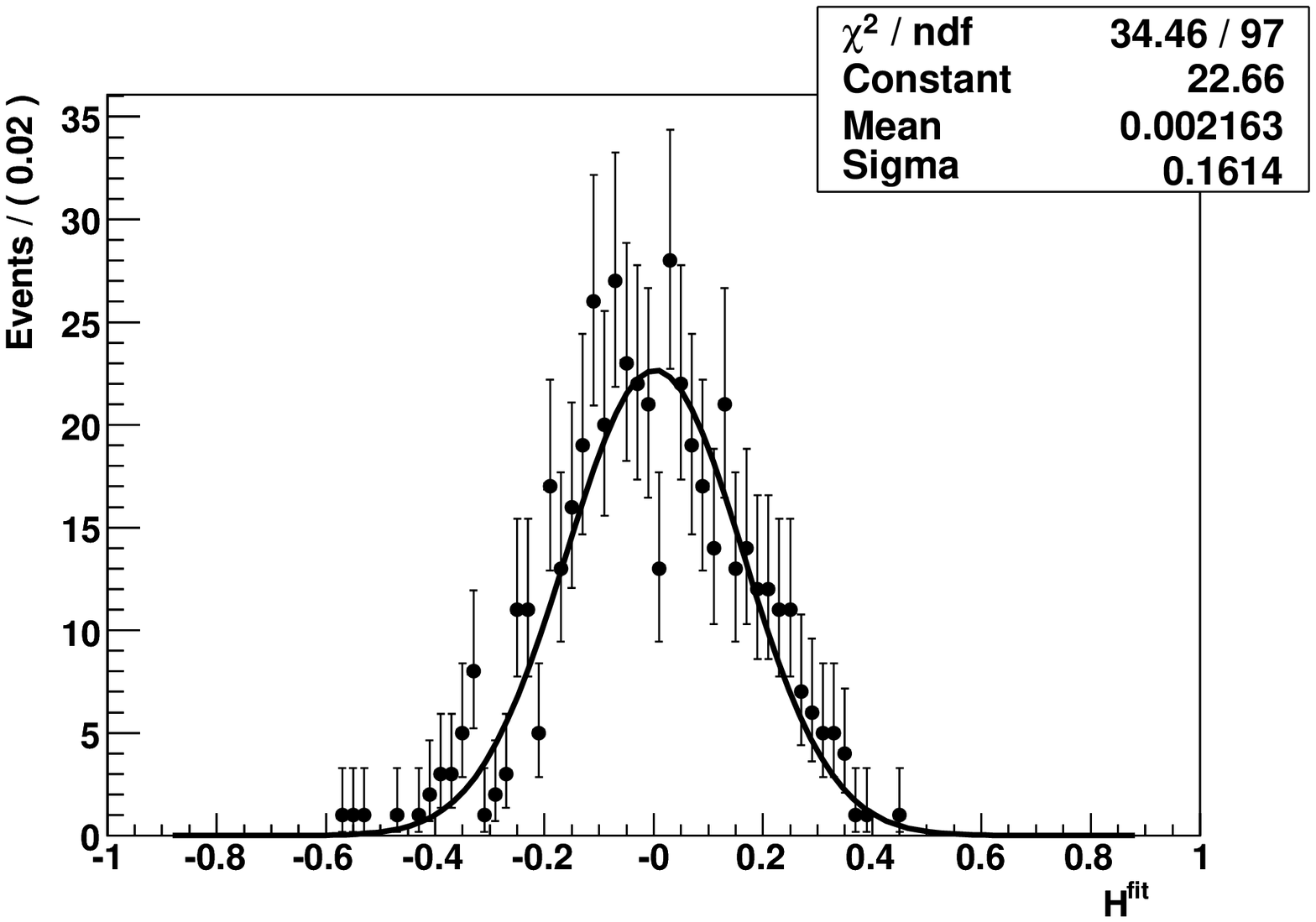}
\includegraphics[width=6.5cm]{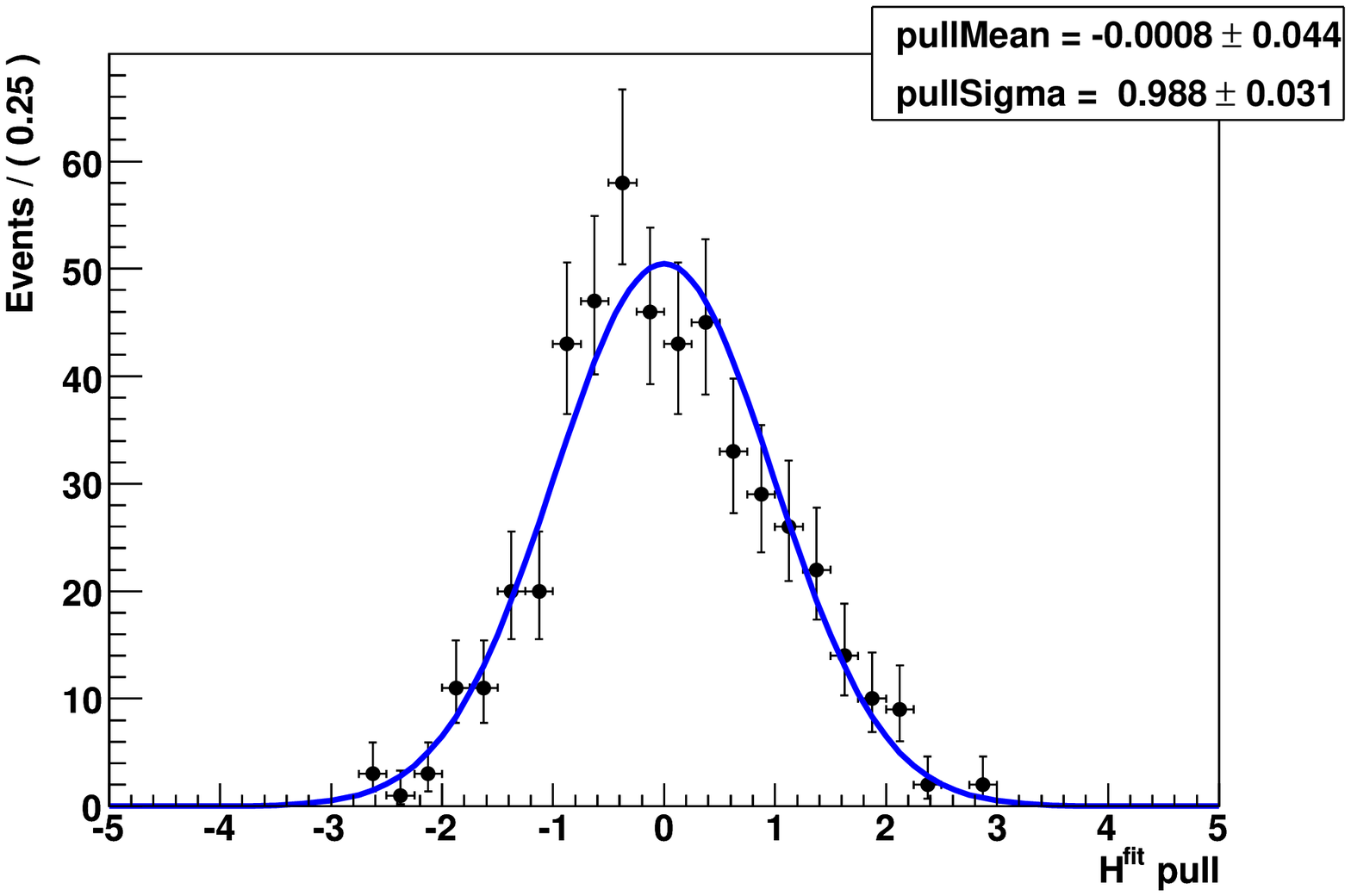}
 \caption{
  Left: the distribution of the fitted values $H^{fit}$ from 500 toy experiments
  for the baseline scenario with $H^{input} =0$ ;
  right: the distribution of $(H^{fit}-H^{input})/\epsilon_H$.
 }
\label{fig:dreso}
\end{center}
\end{figure}

Fig.~\ref{fig:sreso} (left) shows the distribution of all the fitted $S$ values.
A single Gaussian fit is superposed.  
Fig.~\ref{fig:dreso} (left) shows the distribution of the fitted $H$ values, superposed
is a single Gaussian fit.  
We obtain the following sensitivities which are based on a 2 \invfb\ data sample:
$\sigma_S = 0.14$ for $S$, 
$\sigma_H = 0.16$ for $H$ 
and $\sigma_C = 0.15$ for $C$, respectively.
The pull distributions $(S^{fit}-S^{input})/\epsilon_S$ and 
$(H^{fit}-H^{input})/\epsilon_H$
are shown in Fig.~\ref{fig:sreso} (right) and Fig.~\ref{fig:dreso} (right),
respectively.
Here $\epsilon_S$ and  $\epsilon_H$ denote
the errors of  $S^{fit}$ and $H^{fit}$ obtained from each fit, 
These are consistent with standard normal distributions. 
We have repeated these studies for different values of  $C$, $S$ and $H$.
The results for all expected sensitivities on $C$, $S$ and $H$ are summarized in
Table~\ref{table:csd}. It is apparent from the table that 
these sensitivities depend only very weakly
on  the input values.

\begin{table}[bht]
\begin{center}
\caption{Statistical precision of $S$, 
$H$ and $C$ with 2 \invfb\ of data for 
different input values.}
\label{table:csd}
\vspace{0.3cm}
\begin{tabular}{@{}|c|c|c|c|c|c|}
\hline
 $C$ & $S$ & $H$ & 
$\sigma_S$ & 
$\sigma_H$ & 
$\sigma_C$ \\ \hline 
 0.0 & 0.0 & 0.0 & 0.14       & 0.16       & 0.15       \\ \hline
 0.0 & 0.5 & 0.0 & 0.13       & 0.16       & 0.13       \\ \hline
 0.0 & 0.0 & 0.5 & 0.13       & 0.14       & 0.14       \\ \hline

 0.1 & 0.0 & 0.0 & 0.14       & 0.17       & 0.15       \\ \hline
 0.1 & 0.5 & 0.0 & 0.14       & 0.17       & 0.14       \\ \hline
 0.1 & 0.0 & 0.5 & 0.14       & 0.14       & 0.15       \\ \hline

 0.2 & 0.0 & 0.0 & 0.15       & 0.17       & 0.14       \\ \hline
 0.2 & 0.5 & 0.0 & 0.15       & 0.15       & 0.14       \\ \hline
 0.2 & 0.0 & 0.5 & 0.14       & 0.14       & 0.14       \\ \hline

\end{tabular}
\end{center}
\end{table}

Finally we would like to mention that the results of this 
Monte Carlo simulation were obtained with simplified assumptions 
for the experimental conditions, 
e.g. uniform background distributions and perfectly known detector effects.
This is sufficient for this purpose,
which is to demonstrate the clear potential of the 
decay $B_s \to \phi \gamma$ as a probe for new physics 
and to identify the experimental observables 
that are sensitive to right-handed currents. 
We expect that a more complete and detailed study 
of this channel will be performed which 
give more accurate information 
on the physics reach of  $B_s \to \phi \gamma$ at the LHCb experiment.

\section{Conclusions}
\label{sec:conclusions}

The sizable lifetime difference of the $B_s$ meson allows us to 
measure the photon polarization in the  time dependent decay rate of
$B_s \to \phi \gamma$.
In addition to  measuring the coefficient $S$ of the  $\sin(\Delta m t)$  term, 
which is already probed in
$B_d \to K^{*0}(K_S \pi^0)  \gamma$, 
there exists a 
measurable coefficient $H$  for the $\sinh(\Delta \Gamma/2  t)$ term 
in the decay $B_s \to \phi \gamma$.
Both $S$ and $H$  are sensitive to right-handed currents in 
$B \to V \gamma$ transitions.  

The SM prediction, $S_{\phi \gamma} \sim 0 \pm 0.002 $ and 
$H_{\phi \gamma} = 0.047 \pm0.025+0.015 $ \eqref{eq:result}
\cite{charmloops}, 
is dominated by short distance penguins and under control
due to the smallness of the charm loop contributions \eqref{eq:Lc}
originating from the current-current operator \eqref{eq:Q2U}.
We also give a prediction for the direct CP asymmetry,
$C_{\phi \gamma} = 0.005(5)$ \eqref{eq:Cres}, which is sensitive
to new weak phases rather than right-handed currents.

In section \ref{sec:montecarlo}  
we presented a toy Monte Carlo simulation for the
time dependent decay rate of $B \to \phi \gamma$ for 
a  data sample of $2\; {\rm fb}^{-1}$ which will 
be recorded by the LHCb experiment.
From this study we estimate an experimental sensitivity on $S$ of about $0.14$.
The sensitivity on $H$ is inversely proportional to the \Bs\ 
width difference $\Delta \Gamma_s$.
For an anticipated relative width difference of $\Delta \Gamma_s/\Gamma_s = 0.15$ 
a presicion of $0.16$ can be reached for  the
observable $H$.
Note also that $H$ can be extracted from the untagged decay rate.
Thus knowledge of the production flavour of the $B_s$ meson 
is not required (\ref{eq:b-t},\ref{eq:bbar-t})
which will facilitate this measurement.
If either $S$ or $H$ is large in NP, the LHCb experiment 
will be able to observe it.
It is likely that NP, in terms of right-handed currents, will enhance
 the observable $H$ rather than $S$. 
Therefore it is fortunate that the sizable width difference of the $B_s$ meson
gives access to $H$ which 
makes the decay $B_s \to \phi \gamma$ an exciting channel to
search for NP.

\section*{Acknowledgments}

RZ is grateful to Patricia Ball for collaboration on related work and to Thorsten Feldmann for interesting discussions.
He is supported in part  by the Marie Curie research training networks contract Nos.\
MRTN-CT-2006-035482, {\sc Flavianet}, and MRTN-CT-2006-035505, {\sc Heptools}.

\appendix
\setcounter{equation}{0}
\renewcommand{\theequation}{A.\arabic{equation}}

\section{Appendix}
\label{app:formulas}
In this appendix we shall derive the CP asymmetries in terms
of two amplitudes, of different weak and strong phases.
The algebra  
can easily be generalised to an arbitrary number of amplitudes.
We extend  the shorthand notation of Eq.~\eqref{eq:para} to
\begin{equation}
\bar {\cal A}_{L(R)} \equiv {\cal A}[ \bar B_s \to \phi \gamma_{L(R)}] =
 \sum_{i} A^i_{L(R)} e^{i\delta^i_{L(R)}} e^{i \phi^i_{L(R)}} \,
\end{equation}
where $i$ sums over  the amplitudes. 
The weak phase $\phi$ and the strong phase $\delta$ have been  separated leaving the remaining parameter $A^i_{L(R)}$ real. 
In this notation the 
right handed amplitude and the corresponding CP conjugate amplitudes become
\begin{alignat}{2}
\label{eq:A}
 &\bar {\cal A}_L= \sum_{i} A^i_L e^{i\delta^i_L} e^{i \phi^i_L} \quad \stackrel{CP}{\to}  \quad
 {\cal A}_R &=& \xi \sum_i A^i_L e^{i\delta^i_L} e^{-i \phi^i_L} \nonumber \\
 &\bar {\cal A}_R = \sum_i A^i_R e^{i\delta^i_R} e^{i \phi^i_R} \quad \stackrel{CP}{\to} \quad 
 {\cal A}_L &=& \xi \sum_i A_R^i  e^{i\delta^i_R} e^{-i \phi^i_R} \,, 
\end{alignat}
where $\xi$ is the CP-eigenvalue of the final state $V$ and $i = \{u,c,t\}$ is 
the summation over the up-type quarks.  
For $V = \{\rho,\omega,\phi,K^*(K_S\pi^0)\}$ the eigenvalue is $\xi  = 1$ and
for $V=K^*(K_L\pi^0)$ it is $\xi  = -1$ .

In the SM there are three amplitudes at first, corresponding to
the three up-type quarks $u,c$ and $t$
\begin{equation}
\label{eq:amp_dec}
{\cal A} = {\cal A}^u + {\cal A}^c  +   {\cal A}^t  
= \lambda_u { a}^u +  \lambda_c { a}^c +  \lambda_t { a}^t \, ,
\end{equation}
where we have separated out the CKM parameters
$\lambda_U =  V_{\rm  Us}^* V_{\rm Ub}$. The parameters $a^{u,c,t}$
are the same ones as in Eq.\eqref{eq:dec2} up to the helicity specification 
and an irrelevant normalization factor and differ from $A^{u,c,t}$ \eqref{eq:A} 
by the inclusion 
of the strong phase.
As discussed in the main text the three generation unitarity, $\lambda_u + \lambda_c + 
\lambda_t = 0$, may be used to reduce one amplitude, e.g.
 \begin{equation}
\label{eq:Auc_neq} 
{\cal A} = 
 \lambda_t ({ a}^t - { a}^c) +  \lambda_u ({ a}^u - { a}^c) \, ,
\end{equation}
for the sake of more compact formulae.
In the case where the two amplitudes are degenerate, e.g.  
${ a^u} = { a^c}$ the amplitude reduces to a single term. 
This arises in the decay $B \to V \gamma$ if the operators 
$Q_{2}^{u,c}$ \eqref{eq:Q2U}  are not treated separately.
In terms of two amplitudes denoted by $(t,u)$ the CP asymmetries
\eqref{eq:CPamp} 
assume the following form
\begin{eqnarray}
\label{eq:twoamp}
C &=& \frac{4}{N}(( A_L^t A_L^u 
\sin(\phi^t_L\!-\!\phi^u_L) \sin(\delta^t_L\!-\!\delta^u_L) + \{ L \leftrightarrow R \}) \nonumber  \\
H[S] = &\pm &\xi \frac{4}{N}
\big( A_L^t A_R^t \cos(\delta^t_L\!-\!\delta^t_R)\cos[\sin](\phi_s \!-\! \phi^t_L\!-\!\phi^t_R)   \nonumber  \\[0.1cm]
&+& A_L^t A_R^u \cos(\delta^t_L\!-\!\delta^u_R)\cos[\sin](\phi_s \!-\! \phi^t_L\!-\!\phi^u_R\big)
+ \{ u \leftrightarrow t \}) 
\end{eqnarray}
with the normalisation factor 
\begin{equation}
N =  2\big((A_L^u)^2 + (A_L^t)^2 + 2 A^u_L A^t_L \cos(\delta^t_L\!-\!\delta^u_L) \cos(\phi^t_L\!-\!\phi^u_L)+  \{ L \leftrightarrow R \}]\big).
\end{equation} 
Notice that the quantities $H$ and $S$ differ by a cosine and a sine
of the weak phases only.
In the case where there is only one amplitude the 
direct CP asymmetry $C$ vanishes and the formulae
for $S$ and $H$  reduce to
\begin{equation}
H[S] = \xi \frac{\pm 2 A_L A_R \cos(\delta_L\!-\!\delta_R) \cos[\sin](\phi_s \!-\! \phi_L\!-\!\phi_R)}{(A_L)^2 + (A_R)^2}
\end{equation}
The formula for $S$ reduces to the one given in \cite{AGS97} in the case where 
the strong phases $\delta$ are set to zero.
The celebrated formula for
$S_{B_d \to J/\Psi \,K_S}= \sin(2\beta)$ is obtained by setting 
$A_R = A_L$ and $\xi_{J/\Psi K_S}=-1$.

A few remarks concerning the SM are in order.
The  left and right handed phases are equal,
$\phi_L^{\rm SM} = \phi_R^{\rm SM}$. 
The weak mixing phase, as previously mentioned, is approximately given by the top 
quark box diagram  $\phi_s \simeq 2 {\rm Arg}[\lambda_t]
= -2\lambda^2 \eta \simeq -0.035 \simeq -2^\circ$.

The weak phases of the amplitudes are exactly given by
$\phi^U = {\rm Arg}[\lambda_U]$. 
Using the Wolfenstein parameterisation the SM the phases are:
\begin{alignat}{4}
\label{eq:mixi}
&\phi_s  \simeq -2 \lambda^2 \eta\;, \quad
&\phi^t_{b \to s} & \simeq   -\lambda^2 \eta \;, \quad
&\phi^u_{b \to s} & \simeq  -\gamma \;, \quad 
&\phi^c_{b \to s} & = O(\lambda^6)     \;.
\end{alignat}
In the $B_s$ system the unitarity triangle follows
the hierarchical pattern $|\lambda_c| \simeq |\lambda_t| \gg
|\lambda_u|$. The term proportional to $\lambda_u$ in 
\eqref{eq:Auc_neq}  can be neglected in the case where the
trigonometric function of the angles are of the same order,
which is the case for $H$ but not for $S$.


\end{document}